\documentstyle[12pt]{article}
\def\dfrac{\displaystyle\frac}
\def\i{\imath}
\def\a{\alpha}
\def\sn{{\rm sign}}
\def\d{\partial}
\def\q{QED$_{2+1}$}
\def\noi{\noindent}
\def\ve{\varepsilon}
\def\ce{{\cal E}}
\def\bare{\bar{\cal E}}
\def\br{\bar{\rho}}

\newcommand{\refc}[1]{Ref.~\cite{#1}}

\newcommand{\bea}{\begin{eqnarray}}
\newcommand{\eea}{\end{eqnarray}}
\newcommand{\be}{\begin{equation}}
\newcommand{\ee}{\end{equation}}
\newcommand{\bc}{\begin{center}}
\newcommand{\ec}{\end{center}}
\newcommand{\ba}{\begin{array}}
\newcommand{\ea}{\end{array}}

\newcommand{\cL}{{\cal L}}
\newcommand{\Leff}{\cL^{\rm eff}(B,\mu)}
\newcommand{\Lb}{\cL^{\rm eff}(B)}
\newcommand{\Lm}{\tilde\cL^{\rm eff}(B,\mu)}

\newcommand{\ijmp}[3]{{\it Int. J. Mod. Phys. } {{\bf #1} {(#2)} {#3}}}

\newcommand{\np}[3]{{\it  Nucl. Phys. }{{\bf #1} {(#2)} {#3}}}

\newcommand{\prd}[3]{{\it  Phys. Rev. D} {{\bf #1} {(#2)} {#3}}}

\newcommand{\pl}[3]{{\it  Phys. Lett. }{{\bf #1} {(#2)} {#3}}}

\newcommand{\sovjnp}[3]{{\it Sov. J. Nucl. Phys. }{{\bf #1} {(#2)} {#3}}}

\normalsize

\begin{document}

\large
\thispagestyle{empty}
\begin{flushright}                              FIAN/TD/96-09\\
                                                hep-th/9605194\\
                                                May 1996

\vspace{2cm}
\end{flushright}
\bc
\normalsize
{\LARGE\bf Spontaneous Magnetization in Maxwell QED$_{2+1}$}

\vspace{3ex}

{\Large Vadim Zeitlin$^{\dagger\star}$}

{\large Department of Theoretical Physics, P.~N.~Lebedev Physical
Institute,

  Leninsky prospect 53, 117924 Moscow, Russia}\vspace{5ex}

\ec

\centerline{{\Large\bf Abstract}}

\normalsize

\begin{quote}
Spontaneous magnetization in the Maxwell QED$_{2+1}$ at
finite fermion density is studied. It is shown that at low fermion
densities  the one-loop free energy has its minimum at some
nonvanishing value of the  magnetic field. The magnetization is due to
the asymmetry of the fermion spectrum of the massive QED$_{2+1}$
in an external magnetic field.

\end{quote}

\vfill
\noindent
$^\dagger$ E-mail address: zeitlin@lpi.ac.ru

\medskip
\noindent
$^\star$ Talk presented at the 2nd International Sakharov Conference on
Physics, \\
Moscow, May 20-24, 1996
\newpage

\normalsize

\setcounter{page}{1}

In (2+1)-dimensional quantum electrodynamics (\q) an addition of the Chern-Simons
term $\frac{\theta}4 \ve_{\mu\nu\alpha}F^{\mu\nu}A^\alpha$ to the bare
Lagrangian drastically modifies the theory \cite{DJT}: the gauge field becomes
massive and, due to the presence of the totally antisymmetric tensor
$\ve_{\mu\nu\alpha}$, the electric and magnetic components of the modified
Maxwell equation are mixed up.  As a consequence, in a static uniform magnetic
field $B$ the electric charge $j^0_{cs} =  \theta B$ associated with the
Chern-Simons term is induced.  On the other hand, in a uniform magnetic field
charge connected to fermions, $j^0_f = \frac{e^2B}{4\pi}
(2N+1)$ is induced \cite{LSW90,LSW91} ($N$ is the number of filled Landau
levels ), thus the electric neutrality  condition in Maxwell-Chern-Simons \q
is $j^0_f + j^0_{cs} = 0$.  Recently Hosotani has shown that in \q , with the
Chern-Simons term in the bare Lagrangian a (self-consistent) neutral
configuration with a uniform magnetic field has the energy lesser than that of
the naive vacuum \cite{H93}.  Since the condition $j^0_f + j^0_{cs} = 0$ also
implies masslessness of the gauge field, the spontaneous magnetization in the
neutral Maxwell-Chern-Simons \q has Nambu-Goldstone origin [5-8].

In this talk we shall discuss the possibility of the spontaneous magnetization
in the {\it Maxwell} \q with a finite fermion density.  We
shall consider \q with two-component massive fermions
with the Lagrangian

\medskip
        $$
         \cL = -\frac14 F_{\mu\nu}F^{\mu\nu} +\bar{\psi}(\imath {\partial
        \kern-0.5em/} + e {A\kern-0.5em/}   -m)\psi~~~,
        \eqno{(1)}
         $$

\medskip
\noindent
$\gamma$-matrices are the Pauli matrices, $\gamma^0 = \sigma_3, \gamma^1 =
\i\sigma_1, \gamma^2 = \i\sigma_2$. Let us remind that with a uniform external
magnetic field $B=\d_1A_2-\d_2A_1$ the fermion energy spectrum (Landau levels)
in this theory is the following \cite{C,Z89}:

\medskip
        $$
        p_0^{(0)} = -m~\sn (eB), \quad
        p_0^{(\pm n)} = \pm  \sqrt{m^2 + 2|eB|n}, \quad n= 1,2, \dots
        \eqno{(2)}
        $$

\noi
(note the asymmetry of the spectrum. The degeneracy of all levels is
$\frac{|eB|}{2\pi}$).

Addition of the term (-$\mu\psi^\dagger\psi$) to the Lagrangian (1) provides the
parameter controlling the fermion density ($\mu$ is the chemical potential,
therefore all levels with energies up to $\mu$ are filled). One may calculate
the fermion density (induced charge) in \q ~with $B,\mu \ne 0$ using either
the corresponding Green function
\cite{LSW91} or the spectral properties of the theory \cite{N85,Z95}.
The latter is most straightforward way: here, in case of discrete and equally
degenerated levels the density is

\medskip
        $$
        \rho = \dfrac{|eB|}{2\pi}
        \left\{
        - \frac12 \sum_n \sn (p_0^{(n)}) +
        \sum_n
        \left(
        \theta(p_0^{(n)})  \theta(\mu - p_0^{(n)})  -
        \theta(- p_0^{(n)})  \theta(p_0^{(n)}  - \mu)
                                        \right) \right\},
        \eqno{(3)}
        $$

\noi
or,

        $$
        \rho (B,\mu) = \frac{eB}{4\pi} +
        \left\{
        \begin{array}{cl}
        {}~~\dfrac{|eB|}{2\pi}
        \left(\left[
        \dfrac{\mu^2 -m^2}{2|eB|} \right]
        + \theta(-eB)                   \right),\quad     & \mu>\/m;
        \\
        &\\
        0,                   \quad             & |\mu|<m;\\
        &\\
        {}-
        \dfrac{|eB|}{2\pi}
        \left(
        \left[
        \dfrac{\mu^2 - m^2}{2|eB|}      \right]
        + \theta(eB)        \right) ,\quad &\mu<- m
        \end{array}\right.
        \eqno{(4)}
        $$

\medskip
\noindent
($[ \dots ]$ denotes the integral part). The asymmetry of the density as the
function of  $\mu$ is the consequence of the asymmetry of the fermion spectrum,
 $\left[ \frac{\mu^2 -m^2}{2|eB|} \right]$ or ($\left[ \frac{\mu^2
-m^2}{2|eB|} \right] +1$) describes the number of filled Landau levels.
Unlike QED$_{3+1}$  \cite{PZ} the
magnitude of the magnetic field and the number of filled Landau levels do not
define the chemical potential unambiguously -- at a fixed $B$ the fermion
density is a step-function of the chemical potential, Fig. 1.

It follows from Eq.(4) that the equation $\rho(B,\mu) = {\rm const}$ has
an infinite number of solutions enumerated by the filled Landau
levels number \cite{Z95}. For instance, at $\rho= {\rm const}$, ~$\mu, B > 0$
the solutions are:

        $$
        eB = \dfrac{4\pi\rho}{2N+1}, \quad N=0,1,2, \dots
        \eqno{(5)}
        $$

\noi
In $(B,\mu)$--plane these solutions are intervals parallel to the $\mu$--axis,
Fig. 2 (at the vanishing magnetic field  the fermion density is $\rho(\mu) =
\frac1{4\pi}(\mu^2-m^2)\theta(\mu^2-m^2)\sn (\mu) $).

One may minimize the energy of the above-mentioned configurations of the
equal fermion density varying  magnetic field. The energy density is
$\ce = \mu\rho - \Leff = \mu\rho +\frac{B^2}2 - \Lb - \Lm$. The one-loop
effective Lagrangian $\Lb$ was calculated in \refc{R}:

        $$
        \Lb = \dfrac1{8\pi^{3/2}} \int_0^\infty \dfrac{d\!s}{s^{5/2}}
        e^{-m^2s} (eBs \coth(eBs) -1)~~~,
        \eqno{(6)}
        $$

\noi
while the $\mu$--dependent contribution $\Lm = {\displaystyle \int_0^\mu}\rho(B,
\mu')d\!\mu'$ may be easily obtained from Eq.(4).

We shall start with $\mu, B >0$ case and take  $eB \ll m^2$. The energy
density $\ce_N$ of the configuration with specific $N$ is:

        $$
        \ce_N = \dfrac{B^2}2
        \left(  1 - \dfrac{e^2}{12\pi m}        \right)
        + \dfrac{eB}{2\pi} \sum_{n=1}^N \sqrt{m^2 + 2eBn}
        \eqno{(7)}
        $$

\noi
(the second term in the right-hand side of Eq.(7) is just the sum of the
energies of the filled Landau levels $\times$ degeneracy).

The energy density $\ce_N$ may be rewritten in terms of the fermion density,

        $$
        \ce_N = \dfrac{8\pi^2\rho^2}{e^2(2N+1)^2}
        \left(  1 - \dfrac{e^2}{12\pi m}        \right)
        + \dfrac{2\rho}{2N+1}
        \sum_{n=1}^N \sqrt{m^2 + \dfrac{8\pi\rho n}{2N+1}}
        \eqno{(8)}
        $$

\noi
or, by introducing dimensionless variables $\bare_N = \ce_N/m^3, \br=\rho/m^2$
and $\alpha = e^2/8\pi m$,

        $$
        \bare_N = \dfrac{\br^2}{\a(2N+1)^2}
        \left(  1 - \frac23 \a        \right)
        + \dfrac{2\br}{2N+1}
        \sum_{n=1}^N \sqrt{1 + \dfrac{8\pi\br n}{2N+1}}~~~.
        \eqno{(9)}
        $$

To analyze the possibility of the spontaneous magnetization one must compare
the energy $\ce_N$ of the configuration with $N$ filled Landau levels with the
energy $\ce_\mu$ of the configuration of the same density but zero magnetic field.
  If for some $N$, one has $\ce_N < \ce_\mu$, then the spontaneous
magnetization may take place. The energy density $\ce_\mu$ is equal to $\ce_\mu
= \frac1{6\pi} (\mu^3 - m^3)\theta(\mu^2-m^2)$, or $\bare_\mu =
\frac1{6\pi}((1+4\pi\br)^{3/2} -1)$, thus the condition of magnetization is
as follows:

        $$
        \dfrac1{6\pi}\left( (1+4\pi\br)^{3/2} - 1 \right) >
        \dfrac{\pi\br^2}{\alpha(2N+1)^2} (1-{2\over 3}\alpha)
        + \dfrac{2\br}{2N+1}
        \sum_{n=1}^N \sqrt{1+\dfrac{8\pi\br n}{2N+1}}~~~.
        \eqno{(10)}
        $$

   Below we shall suppose $\br$ to be small,  $\br \ll 1$. After making an
expansion in powers of $\br$ in the inequality (10) one has

        $$
        \dfrac1{6\pi} (\frac32 4\pi\br + {3\over 8} (4\pi\br)^2) >
        \dfrac{\pi\br^2}{\a (2N+1)^2} (1-{2\over 3}\a)
        + \dfrac{2\br}{2N+1}
        \sum_{n=1}^N
        \left(   1 + \dfrac{4\pi\br n}{2N+1}        \right)+{\cal O}(\br^3)~~~.
        \eqno{(11)}
        $$

\noi
By collecting similar terms, we finally obtain:

        $$
        \dfrac{\br}{2N+1} >
        \dfrac{\pi\br^2}{\a (2N+1)^2}
         - \dfrac53 \dfrac{\pi\br^2}{(2N+1)^2} +
        {\cal O}(\br^3) ~~~,
        \eqno{(12)}
        $$

\noi
i.e. inequality (10) does have solutions for $\br < \dfrac{\alpha}{\pi}(2N+1)$.

        Thus we have shown that the spontaneous magnetization may take place in
the Maxwell \q~ at low fermion density: at $\br <
\frac{\alpha}{\pi}(2N+1)\ll1$ the energy is minimum when
$\left[\frac{\pi\rho}{\a}\right]$ or ($\left[\frac{\pi\rho}{\a}\right] +1$)
Landau levels are filled. For example, at $\rho < \frac{e^2m}{16\pi}$ the
induced magnetic field is $B=\frac{4\pi\rho}{e}$.

Let us check now whether magnetization is possible with $\mu >0, ~B< 0$. In
this case the fermion density is $\rho = \frac{|eB|}{4\pi} (2N+1)$, $N=1,2,
\dots$,

        $$
        \ce_N = \dfrac{8\pi^2\rho^2}{e^2(2N+1)^2}
        \left(
        1- \frac{e^2}{12\pi m}                  \right)
        +\dfrac{2\rho}{2N+1}
        \sum_{n=0}^N \sqrt{m^2+\dfrac{8\pi\rho n}{2N+1}}
        \eqno{(14)}
        $$

\noi
and the magnetization condition $\ce_\mu > \ce_N$ is equivalent to

        $$
        \br + \pi\br^2 > \dfrac{\pi\br^2}{\a(2N+1)^2}
        + \dfrac{2N+2}{2N+1} \br +\dfrac{4(N+1)(N+2)}{(2N+1)^2}\pi\br^2
        +{\cal O}(\br^3)~~~,
        \eqno{(15)}
        $$

\noi
which has no solutions.

Therefore, for $\mu > 0$, i.e. for finite {\it particle} density the
spontaneous magnetization arises with a definite sign of magnetic field ($B>0$).
For $\mu < 0$, i.e. for finite {\it antiparticle} density the field has the
opposite sign ($B<0$).

To better understand the mechanism of the magnetization it is instructive to
consider (2+1)-dimensional field theory possessing symmetrical fermion spectrum
in a uniform magnetic field, e.g.  \q with four-component spinors. Calculating
$\ce_\mu$ and $\ce_N$ in this model one may see that the energy of magnetized
system is always greater than that with $B=0$ (as well as in QED$_{3+1}$).

\medskip
The examples described above indicate that spontaneous magnetization in the
finite fermion density \q ~is due to the asymmetry of the fermion spectrum  in
a magnetic field. The latter is of topological
nature (see \refc{N85} and references therein).  Therefore, the above arguments
for magnetization based on the one-loop calculations will hold true in all
orders of the perturbation theory.


\bigskip
\section*{Acknowledgments}
This work was supported in part by RBRF grants $N^o$
96-02-16287-a and 96-02-16117-a.


\newpage
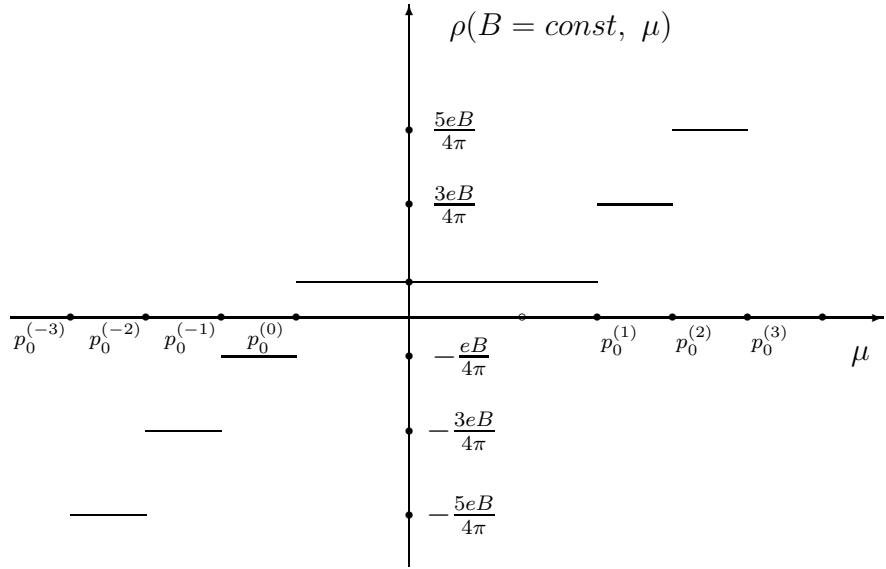
\begin{figure}
\unitlength=1.00mm
\linethickness{0.4pt}
\begin{picture}(143.00,111.83)
\put(55.00,70.11) {\makebox(0,0)[cc]{{\tiny $\bullet$}}}
\put(45.00,70.11) {\makebox(0,0)[cc]{{\tiny $\bullet$}}}
\put(35.00,70.11) {\makebox(0,0)[cc]{{\tiny $\bullet$}}}
\put(105.00,70.11){\makebox(0,0)[cc]{{\tiny $\bullet$}}}
\put(115.00,70.11){\makebox(0,0)[cc]{{\tiny $\bullet$}}}
\put(125.00,70.11){\makebox(0,0)[cc]{{\tiny $\bullet$}}}
\put(135.00,70.11){\makebox(0,0)[cc]{{\tiny $\bullet$}}}
\put(80.00,36.99){\vector(0,1){74.84}}
\put(27.00,70.11){\vector(1,0){116.00}}
\put(140.00,64.95){\makebox(0,0)[cc]{$\mu$}}
\put(115.00,108.82){\makebox(0,0)[rc]{$\rho(B=const,~\mu)$}}
\put(108.00,67.10){\makebox(0,0)[cc]{{\scriptsize $p^{(1)}_0$}}}
\put(118.00,67.10){\makebox(0,0)[cc]{{\scriptsize $p^{(2)}_0$}}}
\put(128.00,67.10){\makebox(0,0)[cc]{{\scriptsize $p^{(3)}_0$}}}
\put(51.00,67.50){\makebox(0,0)[cc]{{\scriptsize $p^{(-1)}_0$}}}
\put(41.00,67.50){\makebox(0,0)[cc]{{\scriptsize $p^{(-2)}_0$}}}
\put(31.00,67.50){\makebox(0,0)[cc]{{\scriptsize $p^{(-3)}_0$}}}
\put(80.00,74.84) {\makebox(0,0)[cc]{{\tiny $\bullet$}}}
\put(80.00,85.16) {\makebox(0,0)[cc]{{\tiny $\bullet$}}}
\put(80.00,95.05) {\makebox(0,0)[cc]{{\tiny $\bullet$}}}
\put(80.00,55.05) {\makebox(0,0)[cc]{{\tiny $\bullet$}}}
\put(80.00,43.87) {\makebox(0,0)[cc]{{\tiny $\bullet$}}}
\put(86.00,95.05){\makebox(0,0)[cc]{$\frac{5eB}{4\pi}$}}
\put(86.00,85.16){\makebox(0,0)[cc]{$\frac{3eB}{4\pi}$}}
\put(87.00,55.05){\makebox(0,0)[cc]{$-\frac{3eB}{4\pi}$}}
\put(87.00,43.87){\makebox(0,0)[cc]{$-\frac{5eB}{4\pi}$}}
\put(65.00,70.11) {\makebox(0,0)[cc]{{\tiny $\bullet$}}}
\put(80.00,64.95) {\makebox(0,0)[cc]{{\tiny $\bullet$}}}
\put(95.00,70.11) {\makebox(0,0)[cc]{{\tiny $\circ$}}}
\put(61.00,67.50){\makebox(0,0)[cc]{{\scriptsize $p^{(0)}_0$}}}
\put(65.00,74.84){\line(1,0){40}}
\put(105.00,85.16){\line(1,0){10}}
\put(115.00,95.05){\line(1,0){10}}
\put(55.00,64.95){\line(1,0){10}}
\put(45.00,55.05){\line(1,0){10}}
\put(35.00,43.87){\line(1,0){10}}
\put(87.00,64.95){\makebox(0,0)[cc]{$-\frac{eB}{4\pi}$}}
\end{picture}

\vspace{2cm}
\caption{Fermion density as a function of chemical potential, $B$=const.}

 \end{figure}

\newpage
\begin{figure}
\unitlength=1.00mm
\linethickness{0.4pt}

\vspace{-2cm}
\begin{picture}(145.00,125.00)
\put(20.00,20.00){\vector(0,1){85.00}}
\put(10.00,30.00){\vector(1,0){134.00}}
\put(110.00,30.00){\line(0,1){45}}
\put(50.00,45.00){\line(0,1){30}}
\put(38.00,57.00){\line(0,1){18}}
\put(33.00,62.50){\line(0,1){12.50}}
\put(30.00,65.500){\line(0,1){9.50}}
\put(141.00,22.00){\makebox(0,0)[cc]{$B$}}
\put(12.00,99.00){\makebox(0,0)[cc]{$\mu^2 - m^2$}}
\put(12.00,75.00){\makebox(0,0)[cc]{$4\pi\rho$}}
\put(117.00,23.00){\makebox(0,0)[cc]{\scriptsize{${\rm B}_0$}}}
\put(57.00,22.00){\makebox(0,0)[cc]{$\frac{{\rm B}_0}{3}$}}
\put(42.00,22.00){\makebox(0,0)[cc]{$\frac{{\rm B}_0}{5}$}}
\put(35.00,22.00){\makebox(0,0)[cc]{$\frac{{\rm B}_0}{7}$}}
\put(28.00,22.00){\makebox(0,0)[cc]{$\frac{{\rm B}_0}{9}$}}
\put(110.00,30.00){\makebox(0,0)[cc]{{\tiny $\bullet$}}}
\put(50.00,30.00) {\makebox(0,0)[cc]{{\tiny $\bullet$}}}
\put(38.00,30.00) {\makebox(0,0)[cc]{{\tiny $\bullet$}}}
\put(33.00,30.00) {\makebox(0,0)[cc]{{\tiny $\bullet$}}}
\put(30.00,30.00) {\makebox(0,0)[cc]{{\tiny $\bullet$}}}
\end{picture}

\vspace{2.cm}
\caption{Solutions of the equation $\rho(B,\mu)= \rho$
in $(B,\mu^2 - m^2)$--plane, ~~$eB_0 = 4\pi\rho$}
\end{figure}
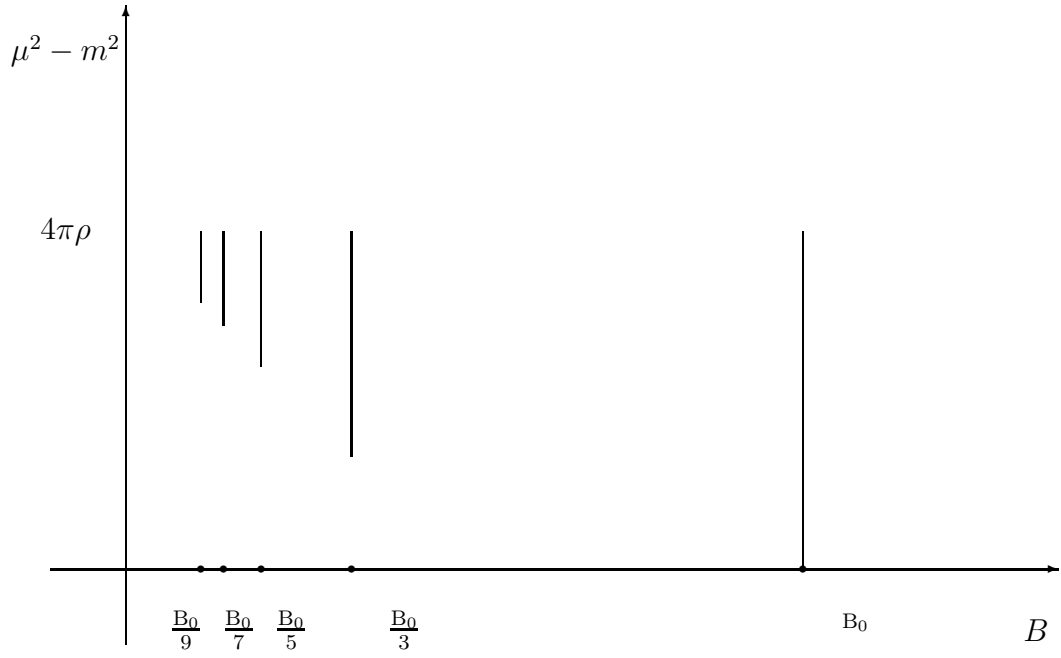

\end{document}